\documentclass[aps,prl,twocolumn,groupedaddress]{revtex4}
\usepackage{amsmath}
\usepackage{graphicx}
\usepackage{color}

\begin{document}

\title{Exciton spectrum in two-dimensional transition metal dichalcogenides:\\
The role of Diracness}

\author{Maxim Trushin$^1$}
\author{Mark Oliver Goerbig$^2$}
\author{Wolfgang Belzig$^1$}
\affiliation{$^1$University of Konstanz, Fachbereich Physik, M703 D-78457 Konstanz, Germany}
\affiliation{$^2$Laboratoire de Physique des Solides, Univ. Paris-Sud, Universit\'e Paris-Saclay, CNRS UMR 8502, F-91405 Orsay, France}

\date{\today}

\begin{abstract}
The physics of excitons, electron-hole pairs that are bound together by their mutual Coulomb attraction, can to great extent be understood in the framework of the quantum-mechanical hydrogen
model. This model has recently been challenged by spectroscopic
measurements on two-dimensional transition-metal dichalchogenides that unveil strong deviations from a hydrogenic spectrum. Here, we show that this deviation is due to the particular
relativistic character of electrons in this class of materials. Indeed, their electrons are no longer described in terms of a Schr\"odinger 
but a massive Dirac equation that intimately links electrons to holes. Dirac excitons therefore inherit a relativistic quantum spin-1/2 
that contributes to the angular momentum and thus the exciton spectrum. Most saliently, the level spacing is strongly reduced as compared to the hydrogen model, in agreement with
spectroscopic measurements and {\it ab-initio} calculations.
\end{abstract}

\maketitle

\section{Introduction}

In this paper, we focus on the problem of excitonic spectra in two dimensional (2D) semiconducting transition metal dichalcogenides (2DTMDs).
An exciton is a bound state of an electron and a hole which are attracted to each other by the Coulomb force \cite{Elliott1957}.
The conventional 2D hydrogen-like exciton spectrum is given by 
\begin{equation}
\label{Rydberg}
 E_{nm}=\Delta -\frac{e^4 \mu}{2\epsilon^2 \hbar^2} \frac{1}{(n+|m|+1/2)^2},
\end{equation}
where $e$ is the elementary charge, $\epsilon$ is the dielectric constant, $\hbar$ is the Planck constant, $\Delta$ is the bandgap, and 
$\mu^{-1}=m_e^{-1}+ m_h^{-1}$ is the reduced mass with $m_{e,h}$ being the electron/hole mass,
and $n=0,1,2...$, $m=0,\pm 1, \pm 2$ are the radial and magnetic quantum numbers, respectively.
However, the exciton spectrum  recently observed in 2DTMDs \cite{PRL2014chernikov,PRL2014he,NanoLett2015hill} does not resemble the conventional Rydberg series (\ref{Rydberg}).
 A few very recent Letters \cite{PRL2016olsen,PRL2015zhou,PRL2015srivastava}
propose different explanations of the nonhydrogenic exciton spectra based on the Berry's phase and non-local screening.
Multiple ab-initio \cite{PRL2013qiu,PRB2012komsa,PRB2013shi,PRB2012cheiwchanchamnangij} and other numerical calculations
\cite{PRB2015wu} have managed to reproduce the non-hydrogenic spectrum, but understanding is still missing from our point of view.
We claim that the main mechanism responsible for the nonhydrogenic Rydberg series in 2DTMDs was overlooked until now.
Here, we develop a new model which takes into account additional angular momentum (aka {\em Diracness}) coming from the pseudospin degree of freedom 
unavoidable in 2D semiconducting materials with honeycomb structure.
The pseudospin changes the magnetic quantum number $m$ in Eq.~(\ref{Rydberg}) to the total angular momentum $j=m+1/2$
and the spectrum is then given by Eq.~(\ref{main}).
The model allows for a transparent interpretation of existing experimental and theoretical data in simple terms, hence, providing understanding that we are striving for.

\section{Model}

The one-particle Hamiltonian for carriers in 2DTMDs is given by \cite{PRL2012xiao},
\begin{equation}
\label{1p}
H_1 = \left(\begin{array}{cc} \Delta/2  & \hbar v k \mathrm{e}^{-i \theta} \\
\hbar v k \mathrm{e}^{i \theta} & -\Delta/2
\end{array}\right),
\end{equation}
where $\tan\theta=k_y/k_x$, and $v$ is the velocity parameter which can be either measured \cite{Kim2016proc} or calculated \cite{PRL2012xiao}.
Due to the spin-orbit coupling in real 2DTMDs $H_1$ splits into two versions with the bandgaps usually denoted by $\Delta_A$ and $\Delta_B$.
The pseudospin degree of freedom encoded in the matrix structure of Eq.~(\ref{1p}) suggests
the possibility that atypical quantum effects can play a role in bound states (\ref{Rydberg}).
The eigenvalues of (\ref{1p}) are $\pm\sqrt{(\hbar v k)^2+\Delta^2/4}$, which in parabolic approximation
suggest the same effective mass $m_{e,h}=\Delta/(2v^2)$ for electrons and holes.

\begin{figure}
\includegraphics[width=\columnwidth]{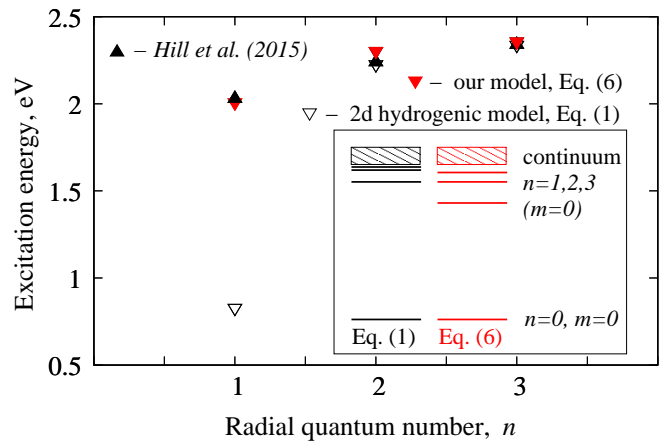}\hspace{2pc}%
\caption{\label{fig1} B-exciton spectrum for 2D MoS$_2$: theory and measurements. 
Only s-states ($m=0$) are optically active.
The measurements for MoS$_2$ are taken from Ref.~\cite{NanoLett2015hill}. The bandgap size $\Delta_B$ is $2.4$eV and
the band parameter $\hbar v = 1.01 \mathrm{eV} \times 3.193$\AA~ \cite{Kim2016proc} result in  $\alpha=0.81$.
The comparison of our model with measurements for WS$_2$ and WSe$_2$ has been done in  \cite{PRB2016trushin}. The inset demonstrates
the relative level spacing in our and conventional models.}
\end{figure}

We assume that the center of mass does not move for optically excited e-h pair \cite{Elliott1957}, and
the electron and hole momenta have the same absolute values but opposite directions.
The two-particle Hamiltonian without Coulomb interactions is therefore given by the tensor product \cite{PRB2013rodin}
$H_2=H_1 \otimes I_2 - I_2 \otimes (T H_1 T^{-1})$ (here $I_2$ is the $2\times 2$ unit matrix, and $T H_1 T^{-1}$ is the time reversal of $H_1$), and reads
\begin{equation}
\label{2p}
H_2=\left( \begin{array}{cccc}
 0 & \hbar k v \mathrm{e}^{i \theta }  & \hbar k v \mathrm{e}^{-i \theta }  & 0 \\
\hbar k v \mathrm{e}^{-i \theta }  & \Delta & 0 & \hbar k v \mathrm{e}^{-i \theta }  \\
 \hbar k v \mathrm{e}^{i \theta }  & 0 & -\Delta & \hbar k v \mathrm{e}^{i \theta }  \\
 0 & \hbar k v \mathrm{e}^{i \theta }  & \hbar k v \mathrm{e}^{-i \theta }  & 0 \\
\end{array}
\right).
\end{equation}
Eq.~(\ref{2p}) can be block-diagonalized into a matrix $H_2=H_2^+\oplus H_2^-$, where
\begin{equation}
\label{H2pm}
H_2^\pm =
\left(\begin{array}{cc}
-\frac{\Delta}{2}\pm\sqrt{\hbar^2 v^2 k^2 +\frac{\Delta^2}{4}} & \hbar v k \mathrm{e}^{i \theta}  \\
 \hbar v k \mathrm{e}^{-i \theta}  &  \frac{\Delta}{2}\pm\sqrt{\hbar^2 v^2 k^2 +\frac{\Delta^2}{4}}  \\
\end{array}
\right).
\end{equation}
The matrix $H_2^+$ has the eigenvalues $E_0=0$ and $E_k=2\sqrt{\hbar^2 v^2 k^2 +\Delta^2/4}$ describing the excitonic states with vanishingly weak interaction.
The diagonal terms in (\ref{H2pm}) can be written within the effective mass approximation, but
the matrix remains in the peculiar mixed ``Dirac-Schr\"odinger'' form: The off-diagonal ``Dirac'' terms $ \hbar v k \mathrm{e}^{\pm i \theta}$
couple the ``Schr\"odinger'' states. Our goal is to write an effective-mass Hamiltonian which mimics this feature,
but remains tractable at the analytical level.
To do that we expand the diagonal terms in $H_2^+$ up to $k^2$-terms, switch on the Coulomb interaction $V(r)=-e^2/\epsilon r$, and change the momenta to the corresponding operators.
The resulting Hamiltonian reads
\begin{equation}
\label{mainH}
\hat H= \left(\begin{array}{cc}  \frac{2\hbar^2 v^2 }{\Delta}(\hat k_x^2 + \hat k_y^2) + V(r) &  \sqrt{2}\hbar v (\hat k_x - i \hat k_y) \\
 \sqrt{2}\hbar v (\hat k_x + i \hat k_y)  & \Delta+V(r)
\end{array}\right),
\end{equation}
which contains no pseudo-differential operators, such as $\sqrt{\Delta^2/4 + \hbar^2 v^2 \hat  k^2}$ that we
would have to deal with starting directly from Eq.~(\ref{H2pm}).
Note, that  our model (\ref{mainH}) has not been obtained directly from the original $4\times 4$ model (\ref{2p}) of coupled 2D Dirac fermions --- indeed, the 
block-diagonalizing transformation  depends on the lattice momentum $k$ and therefore does not commute with the potential $V(r)$. 
However, our hypothesis is that the quantum mechanical nature of pseudospin is more important than that of momentum.
That is why this approximation has been employed even though the quantum-mechanical non-commutativity between $r$ and $k$
will lead to corrective terms that could e.g. be treated perturbatively at a later stage. 
The major merit of our model (\ref{mainH}) is to reproduce the relevant excitonic bands while 
retaining the off-diagonal terms whose manifestation we are investigating here.
In order to find the eigenvalues of (\ref{mainH}) we parametrize the interaction by $\alpha=e^2/\sqrt{2} \epsilon \hbar v$
and assume that $\alpha \ll 1$. This limit corresponds to the experimentally relevant regime of shallow bound states near the edge of continuous spectral branch $E_k$.
The spectrum has been derived in \cite{PRB2016trushin} and reads
\begin{equation}
 \label{main}
 E_{nj} = \Delta - \frac{e^4 \mu}{2\epsilon^2 \hbar^2} \frac{1}{(n+|j|+1/2)^2},
\end{equation}
where $j=m+1/2$ is the {\em total} (i.e. orbital and pseudospin \cite{PRL2012xiao}) angular momentum.
Here, the binding energy is $E_b=e^4 \mu/(2\epsilon^2 \hbar^2)$ is four times smaller than in the conventional model (\ref{Rydberg}).
As compared with Eq.~(\ref{Rydberg}), see Fig.~\ref{fig1}, this spectrum shows much better agreement
with the measurements \cite{PRL2014chernikov,PRL2014he,NanoLett2015hill} and numerical calculations \cite{PRB2015wu,PRL2013qiu}
and, along with the effective Hamiltonian (\ref{mainH}), represents our main finding.

\section{Discussion and conclusion}

First of all, we would like to emphasize that our model Hamiltonian (\ref{mainH}) cannot be
continuously deformed towards the standard hydrogen-like one because of the matrix structure of the former in which the pseudospin angular momentum is encoded. 
Formally, one can obtain the conventional series (\ref{Rydberg}) out of our model at $\alpha\to \infty$ since the diagonal (i.e. ``Schr\"odinger'') 
part in the Hamiltonian (\ref{mainH}) dominates in this limit. However, the state with $E_0=0$ still remains and can overlap with the deep bound states 
making the whole model invalid.

In Table \ref{tabone}, we compare the exciton spectrum (\ref{main}) with the previous numerical models \cite{PRB2015wu,PRL2013qiu},
which also include {\em Diracness} along with many other effects. 
Our model, however, does not take into account non-Coulomb interactions due to the non-local screening in thin semiconductor films \cite{Keldysh,PRL2014chernikov,PRL2016olsen},
as we consider them less important than {\em Diracness}. The non-local screening makes the dielectric constant
$\epsilon$ dependent on the exciton radius which increases with $n$ \cite{PRL2014chernikov,PRL2016olsen}, whereas
{\em Diracness} modifies the very backbone of the exciton model --- the fundamental $1/(n+1/2)^2$ spectral series.
In our model, the dielectric constant $\epsilon$ is solely determined by the SiO$_2$ substrate,
and the Coulomb interaction strength is simply parametrized by the parameter $\alpha$.
Nevertheless, our model qualitatively agrees with the numerical outcomes \cite{PRB2015wu,PRL2013qiu}, as shown in Table \ref{tabone}.
The quantitative version of our model \cite{unpublished} should treat Eq.~(\ref{mainH}) with the Keldysh potential \cite{Keldysh} as $V(r)$.

We also compare the theoretical data with the exciton transition energies measured in \cite{NanoLett2015hill} for MoS$_2$.
As one can see, all the models overestimate the measured excitation energy of B' peak, but more or less agree with each other. 
The latter confirms that our model captures the main mechanism hidden in the numerical calculations.
Note, however, that the measurements \cite{NanoLett2015hill} allow for at least two different assignments of
the exciton spectral lines to the theoretical Rydberg series. 
Nevertheless, we compare the B-exciton spectrum measured there with the outcomes of our model.
One can see that the Rydberg spectrum (\ref{Rydberg}) overestimates the binding energy, whereas Eq.~(\ref{main}) demonstrates much better fit.

\begin{widetext}
\begin{center}
\begin{table}
\begin{tabular}{*{6}{l}}                        
Reference &$A$  & $A'$  & $B$  & $B'$  &  Comments \\
\hline
Ref. \cite{NanoLett2015hill} &1.87\,eV & not resolved &2.03\,eV & 2.24\,eV  & photoluminescence spectroscopy \\
Ref. \cite{PRB2015wu} &1.93\,eV & 2.13\,eV &2.05\,eV & 2.27\,eV & tight-binding calculations \\ 
Ref. \cite{PRL2013qiu} &1.88\,eV & 2.20\,eV &  2.02\,eV & 2.32\,eV & first-principles calculations \\
Eq.~(\ref{main}) &1.87\,eV & 2.14\,eV &2.01\,eV & 2.30\,eV & $\Delta_A=2.24$ eV, $\Delta_B=2.4$ eV \\ 
\end{tabular}
\caption{\label{tabone}Optical transition energies for absorption peaks $A$ (ground state of the exciton $A$), 
$A'$ (its 1st excited state), $B$ (ground state of the exciton $B$), and $B'$ (its 1st excited s-state) obtained from different studies.
Our model employs parameters from \cite{Kim2016proc}: $\hbar v = 1.01 \mathrm{eV} \times 3.193$\AA,
$\Delta_\mathrm{SO}=\Delta_B-\Delta_A=0.16$ eV.  The environment is SiO$_2$ with the dielectric constant $\epsilon=3.9$.} 
\end{table}
\end{center}
\end{widetext}

To conclude, the pseudospin angular momentum (i.e. {\em Diracness}) does not lead just to a correction of the hydrogen-like exciton spectra but qualitatively modifies 
the excitonic Hamiltonian which suggests lower binding energy and reduced level spacing.

\section*{References}

\bibliography{bib-proceedings.bib}

\end{document}